\documentclass[aps,prl,reprint, showpacs,showkeys]{revtex4-1}
\usepackage{braket}
\usepackage{blindtext}
\usepackage{amsmath}
\usepackage{xcolor}
\usepackage{tikz}
\usepackage[pdftex]{hyperref}
\definecolor{purple}{rgb}{1,0,1}
\definecolor{lime}{HTML}{A6CE39} 

\newcommand{\orcidicon}{%
	\begin{tikzpicture}
	\draw[lime, fill=lime] (0,0) 
		circle [radius=0.16] 
		node[white] {{\fontfamily{qag}\selectfont \tiny ID}};
	\draw[white, fill=white] (-0.0625,0.095) 
		circle [radius=0.007];
	\end{tikzpicture}
	\hspace{-3mm}
}
\newcommand\orcidDel{{\href{https://orcid.org/0000-0003-4158-202X}{\orcidicon}}}
\newcommand\orcidMatt{{\href{https://orcid.org/0000-0003-1088-6485}{\orcidicon}}}
\begin{document}
\title{Quantum PBR Theorem as a Monty Hall Game}
\author{Del Rajan\orcidDel\hspace{-2mm}}
\author{Matt Visser\orcidMatt\hspace{-2mm}}

\affiliation{School of Mathematics and Statistics, Victoria University of Wellington, Wellington 6140, New Zealand.\vspace{2pt}}
\date{\LaTeX-ed \today}


\begin{abstract}
The quantum Pusey--Barrett--Rudolph (PBR) theorem addresses the question of whether the quantum state corresponds to a $\psi$-ontic model (system's physical state) or to a $\psi$-epistemic model (observer's knowledge about the system).  We reformulate the PBR theorem as a Monty Hall game, and show that winning probabilities, for switching doors in the game, depend whether it is a $\psi$-ontic or $\psi$-epistemic game. For certain cases of the latter, switching doors provides no advantage.  We also apply the concepts involved to quantum teleportation, in particular for improving reliability. 
\end{abstract}

\maketitle


\textit{Introduction:} No-go theorems in quantum foundations are vitally important for our understanding of quantum physics.  Bell's theorem~\cite{bell1964einstein} exemplifies this by showing that locally realistic models must contradict the experimental predictions of quantum theory.  

There are various ways of viewing Bell's theorem through the framework of game theory~\cite{brunner2014bell}.  These are commonly referred to as nonlocal games, and the best known example is the CHSH game; in this scenario the participants can win the game at a higher probability with quantum resources, as opposed to having access to only classical resources.  There has also been work on the relationship between Bell's theorem and Bayesian game theory~\cite{brunner2013connection, roy2016nonlocal, banik2019two}; in a subset of cases it was shown that quantum resources provide an advantage, and lead to quantum Nash equilibria.  In~\cite{pappa2015nonlocality}, it was shown that quantum nonlocality can outperform classical strategies in games where participants have conflicting interests.  In~\cite{almeida2010guess}, a nonlocal game was constructed where quantum resources did not offer an advantage.

Beyond Bell's theorem, entropic uncertainty relations can be viewed in the framework of a guessing game~\cite{coles2017entropic, coles2019entropic}; the uncertainty relation constraints the participant's ability to win the game.  More broadly, the relationship between quantum theory and game theory is investigated in~\cite{eisert1999quantum, benjamin2001multiplayer, khan2018quantum}. The Monty Hall game~\cite{rodriguez2018probability,rosenthal2008monty, gill2010monty, lucas2009monty} has also been generalized into quantum versions~\cite{li2001quantum,flitney2002quantum,d2002quantum,khan2010quantum,kurzyk2016quantum,zander2006positive,paul2019playing}.

The Pusey--Barrett--Rudolph (PBR) theorem~\cite{pusey2012reality} is a relatively recent no-go theorem in quantum foundations.  It addresses the question of whether the quantum state corresponds to a $\psi$-ontic model (physical state of a system) or to a $\psi$-epistemic model (observer's knowledge about the system)~\cite{harrigan2010einstein}. Notable developments on the PBR theorem and $\psi$-epistemic models have been carried out in~\cite{lewis2012distinct, schlosshauer2012implications, aaronson2013psi, patra2013no, schlosshauer2014no, mansfield2016reality, leifer2017time, leifer2014quantum, jennings2016no}, including on the issue of quantum indistinguishability~\cite{leifer2014psi, barrett2014no, branciard2014psi}, as well being interpreted through the language of communication protocols~\cite{montina2012epistemic, montina2015communication}.  

Analogous to the game formulation of Bell's theorem, a desirable construction is to view the PBR theorem through the lens of a game.  One instantiation of this is in an exclusion game where the participant's goal is to produce a particular bit string~\cite{perry2015communication, bandyopadhyay2014conclusive}; this has been shown to be related to the task of quantum bet hedging~\cite{arunachalam2013quantum}.  Furthermore, concepts involved in the PBR proof have been used for a particular guessing game~\cite{myrvold2018psi}.

In this Letter, we reformulate the PBR theorem into a Monty Hall game.  This particular gamification of the theorem highlights that winning probabilities, for switching doors in the game, depend on whether it is a $\psi$-ontic or $\psi$-epistemic game; we also show that in certain $\psi$-epistemic games switching doors provides no advantage.  This may have consequences for an alternative experimental test of the PBR theorem.  Furthermore, we shall also use the concepts involved for modifying quantum teleportation~\cite{bennett1993teleporting,nielsen2010quantum} to view it as a Monty Hall game.  Using these notions, we develop an error-correcting strategy for unreliable teleportation which may be relevant for practical quantum networks.

\textit{PBR theorem:} We provide a rough sketch of the PBR proof~\cite{pusey2012reality}, and highlight crucial outcomes.  Two quantum systems are prepared independently, and each system is prepared in either state $\ket{0}$ or state $\ket{+}=(\ket{0}+\ket{1})/\sqrt{2}$.  This means that the total system is in one of the four possible non-orthogonal quantum states:
\begin{eqnarray}\label{states}
\ket{\Psi_{1}} &= \ket{0} \otimes \ket{0},  \quad \ket{\Psi_{2}} = \ket{0} \otimes \ket{+},
\nonumber
\\
\ket{\Psi_{3}} &= \ket{+} \otimes \ket{0},  \quad \ket{\Psi_{4}} = \ket{+} \otimes \ket{+}.
\end{eqnarray}
The total system is brought together and measured in the following entangled basis:
\begin{eqnarray}\label{outcomes}
\ket{\Phi_{1}} &= \frac{1}{\sqrt{2}}(\ket{0} \otimes \ket{1} + \ket{1} \otimes \ket{0}),
\nonumber
\\ 
\ket{\Phi_{2}} &= \frac{1}{\sqrt{2}}(\ket{0} \otimes \ket{-} + \ket{1} \otimes \ket{+}),
\nonumber
\\
\ket{\Phi_{3}} &= \frac{1}{\sqrt{2}}(\ket{+} \otimes \ket{1} + \ket{-} \otimes \ket{0}),
\nonumber
\\
\ket{\Phi_{4}} &= \frac{1}{\sqrt{2}}(\ket{+} \otimes \ket{-} + \ket{-} \otimes \ket{+}),
\end{eqnarray}
where $\ket{-}=(\ket{0}-\ket{1})/\sqrt{2}$. 

Invoking the Born probabilities, $\lvert \braket{\Phi_{i}|\Psi_{h}}\rvert^{2}$, where $i,h=1,2,3,4$, we have for $i=h$, $\lvert \braket{\Phi_{i}|\Psi_{i}}\rvert^{2} = 0$. This means that for any value $i$, the outcome $\ket{\Phi_{i}}$ \emph{never} occurs when the system is prepared in quantum state $\ket{\Psi_{i}}$.  The PBR proof showed that in $\psi$-epistemic models there is a non-zero probability $q$ (whose value does not need to be specified) that outcome $\ket{\Phi_{i}}$ occurs when state $\ket{\Psi_{i}}$ is prepared, thereby contradicting the predictions of quantum theory; hence one can infer that the quantum state corresponds to a $\psi$-ontic model.

\textit{Classic Monty Hall:} A character named Monty hosts a game show.  There are three closed doors respectively labelled $\{1,2,3\}$.  There is a prize behind one door, and goats behind the remaining two.  The prize door is denoted $A_{i}$ where $i$ takes one of the door labels, and this choice of prize door is made by the producers of the show.  We assume in the game that when a random choice needs to be made, all options are chosen with the same probability. Hence, we have $P(A_{i}) = 1/3$ for all values $i$. The contestant on the show, who doesn't know which door the prize is behind, gets to pick a door; we label this as $B_{j}$ where $j$ takes door labels; given this is a random choice, we have $P(B_{j}|A_{i})= 1/3$, for all values $i,j$.  Next, Monty who knows where the prize is, has to open a goat door, $C_{k}$ where $k$ takes one door labels.  Monty's decision is constrained through the game rule that he can't open the door chosen by the contestant. Hence we have the following conditional probabilities:
\begin{equation}
P(C_{k} \, | \, B_{j} \cap A_{i}) = 
\begin{cases}
\frac{1}{2},& \text{if } i =j \neq k,\\
1,              & \text{if } i \neq j \neq k, \\
0, & \text{otherwise}.
\end{cases}
\end{equation}
Once a goat door is opened, Monty offers the contestant the option to stick with the original choice or switch to the other unopened door.  By sticking, the contestant's probability of opening the prize door is $1/3$.  Counter-intuitively, by switching doors, the probability of winning increases to $2/3$.  This can be seen by computing the non-zero joint probabilities for all events
\begin{equation*}
P(A_{i} \cap B_{j} \cap C_{k})= P(C_{k} \, | \, B_{j} \cap A_{i})P(B_{j}|A_{i})P(A_{i}),
\end{equation*}     
and then summing those values for the events where the contestant would win by switching. This results in
\begin{equation}
P(\text{win if switch}) = \sum_{i \neq j \neq k} P(A_{i} \cap B_{j} \cap C_{k}) = \frac{2}{3}.
\end{equation}

\textit{Ignorant Monty Hall:} Just as in the Classic case, we have $P(A_{i}) = 1/3$ and $P(B_{j}|A_{i})= 1/3$, for all values $i,j$.  But in this game, Monty doesn't know what lies behind any of the doors.  The only constraint is that Monty can't open the door chosen by the contestant, hence we have   
\begin{equation}
P(C_{k} \, | \, B_{j} \cap A_{i}) = 
\begin{cases}
0,& \text{if } j=k, \\
\frac{1}{2}, & \text{otherwise},
\end{cases}
\end{equation}
There is now a probability that he will open up the prize door by accident, and thus ending the game:
\begin{equation*}
P(\text{opens prize door}) = \sum_{i = k \neq j} P(A_{i} \cap B_{j} \cap C_{k}) = \frac{1}{3}.
\end{equation*} 
This implies that the probability that he opens a goat door is $2/3$. The joint probability that Monty opens a goat door and the contestant wins by switching doors can be computed to be $1/3$.  
From the last two values, we can calculate the conditional probability
\begin{equation}
P(\text{win if switch} \, | \, \text{opens goat door}) = \frac{1/3}{2/3} = \frac{1}{2}.
\end{equation}
This means if Monty opens a goat door, then the contestant's probability of winning is the same whether the contestant chooses to switch the door or not.

\enlargethispage{10pt}
\textit{\textit{$\psi$-ontic Monty Hall game:}} Antidistinguishability~\cite{leifer2014quantum, caves2002conditions, heinosaari2018antidistinguishability}, where there is a measurement for which each outcome identifies that a specific member of a set of quantum states was definitely \emph{not} prepared, is highlighted in the PBR proof by $\lvert\braket{\Phi_{i}|\Psi_{i}}\rvert^{2} = 0$ for all $i$. We will exploit this to construct our game, which can be thought of as a quantum Ignorant Monty Hall game.

For state $\ket{\Psi_{1}}$ in (\ref{states}), we have
\begin{eqnarray}\label{Born}
\lvert\braket{\Phi_{1}|\Psi_{1}}\rvert^{2} &= 0,  \quad \lvert\braket{\Phi_{2}|\Psi_{1}}\rvert^{2} &= 1/4,
\nonumber
\\
\lvert\braket{\Phi_{3}|\Psi_{1}}\rvert^{2} &= 1/4,  \quad \lvert\braket{\Phi_{4}|\Psi_{1}}\rvert^{2} &= 1/2.
\end{eqnarray}
For the other states in (\ref{states}), the same probability distribution ($0$, $1/4$, $1/4$, $1/2$) occur but across the different outcomes (\ref{outcomes}); hence we will focus our game on $\ket{\Psi_{1}}$, but similar constructions hold for the other states.  

The Monty Hall gamification is as follows:  There are four doors labelled $\{1,2,3,4\}$, and these correspond to the different measurement outcomes listed in (\ref{outcomes}).  The prize door $A_{i}$, where $i$ takes one of the door labels, is the outcome $\ket{\Phi_{i}}$ that the state $\ket{\Psi_{1}}$ collapses to upon measurement.  For a $\psi$-ontic game, through the Born probabilities (\ref{Born}), we have $P(A_{i}) = \lvert\braket{\Phi_{i}|\Psi_{1}}\rvert^{2}$.  
  
The contestant on the show doesn't know what state from (\ref{states}) is used, and is only aware of the possible measurement outcomes ($\ref{outcomes}$).  Based on this limited information, the contestant randomly picks one of the doors which we denote $B_{j}$ where $j$ is the corresponding door label; hence we have $P(B_{j}|A_{i})= 1/4$, for all values $i,j$.

Monty's decision corresponds to the predictions of quantum theory.  He is aware that state  $\ket{\Psi_{1}}$ was used, and has access to the Born probabilities (\ref{Born}).  The door opened by Monty is denoted $C_{k}$ where $k$ is one of the door labels.  The main insight to construct this game is that when Monty opens a goat door, he is opening a door that has probability zero of having a prize in it.  And for our game, a door that definitely does not have a prize in it corresponds to outcome $\ket{\Phi_{1}}$ as $P(A_{1}) = \lvert\braket{\Phi_{1}|\Psi_{1}}\rvert^{2} = 0$.  Hence in this game, Monty will open door $C_{1}$ unless the contestant has already chosen this door as their pick (as Monty can't open the door chosen by the contestant); in that case Monty will open one of the other remaining doors with equal probability, and there is a chance he may open up the prize door as in the Ignorant Monty Hall game.  From these factors, one can compute,
\begin{equation}\label{Monty}
P(C_{k} \, | \, B_{j} \cap A_{i}) = 
\begin{cases}
\frac{1}{3},& \text{if } j =1 \text{ and } k=2,3, 4,\\
1,              & \text{if } j \neq 1 \text{ and } k=1, \\
0, & \text{otherwise.}
\end{cases}
\end{equation}
The probability that Monty opens the prize door is
\begin{equation*}
P(\text{opens prize door}) = \sum_{i = k \neq j} P(A_{i} \cap B_{j} \cap C_{k}) = \frac{1}{12}.
\end{equation*}  
This implies that the probability that he opens a goat door is $11/12$.  Monty then offers the option to stick or switch.  Suppose the contestant always sticks with the initial choice.  Then the probability of winning if sticking and Monty opening a goat door is 
\begin{equation*}
\sum_{i = j \neq k} P(A_{i} \cap B_{j} \cap C_{k}) = \frac{1}{4}.
\end{equation*} 
With that, we can compute the conditional probability
\begin{equation}
P(\text{win if stick} \, | \, \text{opens goat door}) = \frac{1/4}{11/12} = \frac{3}{11}.
\end{equation}
Suppose the contestant decides to always switch to one of the other two unopened doors with equal probability $1/2$.  Let $\ket{\Phi_{l}}$ be the outcome switched to and let $D_{l}$ be the corresponding door.  With that, we can compute $P(A_{i} \cap B_{j} \cap C_{k} \cap D_{j}) =  P(D_{l} \, | C_{k} \cap B_{j} \cap A_{i})P(C_{k} \, | \, B_{j} \cap A_{i})P(B_{j}|A_{i})P(A_{i})$.  Hence, the probability of winning if switching and Monty opening a goat door is       
\begin{equation}
\sum_{i = l \neq j \neq k} P(A_{i} \cap B_{j} \cap C_{k} \cap D_{j}) = \frac{1}{3}.
\end{equation}
From that, one can calculate
\begin{equation}
P(\text{win if switch} \, | \, \text{opens goat door}) = \frac{1/3}{11/12} = \frac{4}{11}.
\end{equation}
In a $\psi$-ontic game, switching provides an advantage.

\textit{\textit{$\psi$-epistemic Monty Hall game:}} In the PBR proof, for the $\psi$-epistemic model, there is a non-zero probability $q$ that outcome $\ket{\Phi_{1}}$ occurs when state $\ket{\Psi_{1}}$ is prepared.  This implies that in a $\psi$-epistemic game, $P(A_{1}) = q \neq 0$.  To allow for a comparison with the $\psi$-ontic game, let $q = q_{1} + q_{2} + q_{3}$, and with that let the other prize door probabilities take values $P(A_{2}) = (1/4) - q_{1}$, $P(A_{3}) = (1/4) - q_{2}$ and $P(A_{4}) = (1/2) - q_{3}$.  

As in the $\psi$-ontic game, $P(B_{j}|A_{i})= 1/4$, for all values $i,j$.  Monty as a character corresponds to the predictions of quantum theory (\ref{Born}); he will assume $C_{1}$ is definitely a goat door since $\lvert\braket{\Phi_{1}|\Psi_{1}}\rvert^{2} = 0$.  This means the probabilities in  (\ref{Monty}) apply in this game as well.  Hence, the probability that Monty opens the prize door
\begin{equation*}
P(\text{opens prize door}) = \sum_{i = k \neq j} P(A_{i} \cap B_{j} \cap C_{k}) = \frac{1}{12} + \frac{2q}{3}.
\end{equation*}  
This implies that the probability that Monty opens a goat door is $(11/12) - (2q/3)$.  The probability of winning if always sticking and that Monty opens a goat door is 
\begin{equation*}
\sum_{i = j \neq k} P(A_{i} \cap B_{j} \cap C_{k}) = \frac{1}{4}.
\end{equation*} 
From this we compute
\begin{equation}
P(\text{win if stick} \, | \, \text{opens goat door}) = \frac{3}{11-8q}.
\end{equation}
If a switching strategy is adopted then:      
\begin{eqnarray}
\sum_{i = l \neq j \neq k} P(A_{i} \cap B_{j} \cap C_{k} \cap D_{j}) = \frac{1}{3} - \frac{q}{3}, \\
P(\text{win if switch} \, | \, \text{opens goat door})  = \frac{4-4q}{11-8q}.
\end{eqnarray}
Thus the  probabilities depend on whether the game is a $\psi$-ontic or $\psi$-epistemic game.  For value $q=1/4$, we can calculate that $P(\text{win if switch} \, | \, \text{opens goat door}) = P(\text{win if stick} \, | \, \text{opens goat door})$; hence for certain $\psi$-epistemic games, switching offers no advantage.

\textit{Experimental implications:}  Comparing a $\psi$-ontic game to a $\psi$-epistemic game, Monty opens the prize door less often.  This corresponds to certain probabilities in the PBR proof being zero; some work on the experimental tests~\cite{pusey2012reality, nigg2015can, miller2013alternative, ringbauer2015measurements, liao2016experimental} of PBR discuss this exact zero probability as an experimental difficulty. Through our game, we provide another viewpoint;  the difference in the probabilities of winning conditioned that a goat door is opened are simply different for the two physical scenarios. This may provide insights to alternative experimental designs to test PBR. 

\textit{Quantum teleportation:} Consider the standard protocol~\cite{nielsen2010quantum}. Alice wants to send some unknown state $\ket{\psi} = \alpha{\ket{0}} + \beta{\ket{1}}$ to Bob.  They each possess a member of the Bell state  $\ket{\beta_{00}} = {1\over\sqrt2}(\ket{00}+\ket{11})$.  The initial state is $\ket{\psi} \otimes \ket{\beta_{00}}$.  Alice applies a CNOT gate to her qubits, followed by a Hadamard gate to her first qubit.  The resulting state can be written as
\begin{eqnarray}
\frac{1}{2} \Big( \ket{00}(\alpha\ket{0} + \beta\ket{1}) + \ket{01} (\alpha \ket{1} + \beta \ket{0})
\nonumber
\\
+ \ket{10}(\alpha\ket{0} - \beta\ket{1}) + \ket{11} (\alpha \ket{1} - \beta \ket{0}) \Big).
\end{eqnarray}
When Alices measures her qubits she gets one of the results on the left in  (\ref{Alice}).  Bob would then apply the corresponding Pauli operator on his qubit to obtain $\ket{\psi}$:    
\begin{eqnarray}\label{Alice}
00&  \rightarrow &\text{Does nothing},
\nonumber
\\
01& \rightarrow &\text{Applies } \sigma_{x} = \ket{0}\bra{1} + \ket{1}\bra{0}, 
\nonumber
\\
10& \rightarrow &\text{Applies }  \sigma_{z} = \ket{0}\bra{0} - \ket{1}\bra{1},
\nonumber
\\
11& \rightarrow &\text{Applies }   \sigma_{z}\sigma_{x}.
\end{eqnarray}
Bob receives the two bits from Alice in (\ref{Alice}) through a classical channel.   
This protocol has been extended to probabilistic cases~\cite{li2000probabilistic, lu2000teleportation, agrawal2002probabilistic} and noisy cases~\cite{fortes2015fighting, fortes2016probabilistic, knoll2014noisy, carlo2003teleportation, kumar2003effect}.

\textit{Monty Hall teleportation:} For our first application, we want to modify the standard teleportation protocol into a Monty Hall game.  Alice can be viewed as Monty, and Bob as the contestant.  The four doors are respectively labelled $(00, 01, 10, 11)$.
This coincides with Alice's possible measurement results in (\ref{Alice}); the prize door is Alice's actual result, whose bits we denote $ab$, and what Bob would need get the desired state $\ket{\psi}$. The contestant's initial choice of door would be equivalent to what Bell state was used at the start of the protocol. In this modification, the contestant is allowed to choose any of the four doors $(00, 01, 10, 11)$, which we denote $xy$.  This event coincides with using Bell state
\begin{equation}
\ket{\beta_{xy}} = \frac{1}{\sqrt{2}}(\ket{0}\ket{y} + (-1)^{x}\ket{1}\ket{\bar{y}}), 
\end{equation}
where $\bar{y}$ is the negation of $y$.
As an example, if the contestant chooses door $01$, then a way to implement this is that Bob applies the operator $(\sigma_{0} \otimes \sigma_{z})\ket{\beta_{00}} = \ket{\beta_{01}}$, and communicates that to Alice; the last step would be analogous to Monty being aware of what door the contestant chooses.
In this modified protocol, the initial state is $\ket{\psi}\ket{\beta_{xy}}$.  After Alice applies a CNOT gate to her qubits followed by a Hadamard gate the resulting state is
\begin{eqnarray}
\frac{1}{2} \Big( \ket{00}(\alpha\ket{y} + \beta(-1)^{x}\ket{\bar{y}}) + \ket{01} (\alpha (-1)^{x} \ket{\bar{y}} + \beta \ket{y})
\nonumber
\\
+ \ket{10}(\alpha\ket{y} - \beta(-1)^{x}\ket{\bar{y}}) + \ket{11} (\alpha (-1)^{x} \ket{\bar{y}} - \beta \ket{y}) \Big).
\nonumber
\end{eqnarray}
At this step, Alice measures her qubits to get her result.  If Alice's result is $ab=xy$, meaning it coincides with the Bell state used $\ket{\beta_{xy}}$, then Bob has to do nothing and he has the desired state $\ket{\psi}$ (the exception is if the initial Bell state used was $\ket{\beta_{11}}$ in which case Bob has to apply operator ($-\sigma_{0}$) to get $\ket{\psi}$ if result is $11$).  This is why the contestant's initial choice relates to the Bell state used. 

In this Monty Hall protocol, Alice sends Bob two bits as in (\ref{Alice}) with the following modification: she chooses two bits denoted $cd$  (ie goat door) that are not $xy$ (ie contestant's initial choice) and are not $ab$ (ie prize door). Should Bob do nothing, or apply one of the possible operators (which depend on what Bell state was used) to get $\ket{\psi}$ ie should the contestant stick or switch?

To answer this, let $B_{xy}$ be the door chosen by contestant.  For this example, assume we use $\ket{\beta_{00}}$, hence $P(B_{00})=1$.  Let $A_{ab}$ be the prize door and due to Born probabilities we have $P(A_{ab})= 1/4$.  Let $C_{cd}$ be the goat door opened by Monty whose probabilities, from the protocol description, work out as:   
\begin{equation}
P(C_{cd} \, | \, B_{00} \cap A_{ab}) = 
\begin{cases}
\frac{1}{3},& \text{if } 00 = ab \neq cd,\\
\frac{1}{2},& \text{if } 00 \neq ab \neq cd, \\
0, & \text{otherwise}.
\end{cases}
\end{equation}
If Bob always does nothing (ie, stick strategy), then 
\begin{equation}
P(\text{win if stick}) = \sum_{ab = 00 \neq cd} P(A_{ab} \cap B_{00} \cap C_{cd}) =  \frac{2}{8}.
\end{equation}
Suppose Bob decides to always apply one of the two operators (ie, switch strategy). 
Then there are one of two possibilities which we denote $ef$ and given its a random choice, each occur with probability $1/2$.  Let $D_{ef}$ represent that door, and $P(\text{win if switch})$ is
\begin{equation} \sum_{ab = ef \neq cd \neq 00} P(A_{ab} \cap B_{00} \cap C_{cd} \cap D_{ef}) =  \frac{3}{8}.
\end{equation}
This means Bob should apply one of the two operators (switch) rather than do nothing (stick) to get state $\ket{\psi}$.

\textit{Unreliable teleportation:} For our second application, consider the standard teleportation protocol with the following unreliability: one of the two bits (either the first or second) Alice sends to Bob in (\ref{Alice}) is received but the other is lost; each event occurs with probability $1/2$.  If the initial Bell state is $\ket{\beta_{00}}$ and Alice's result is $00$, then Bob can do nothing. But in this scenario, if Bob receives the single bit as $1$, then the possible options are $01, 10, \text{or }11$; in this case he should apply one of the operators (switch).  If Bob receives bit $0$, then his options are $00, 01, 10$.  Should he stick (to $00$) or switch (to $01$ or $10$)?  To answer this, let us use the notation developed.

We have $P(B_{00})= 1$ and $P(A_{ab}) = 1/4$ . Let $d$ in $C_{d}$ be the single bit received by Bob; based on the scenario described above, we have $P(C_{0} \, | \, B_{00} \cap A_{00}) = 1,$ $P(C_{0} \, | \, B_{00} \cap A_{01}) = 1/2$, and $P(C_{0} \, | \, B_{00} \cap A_{10}) = 1/2$.  We can compute the probability that Bob receives bit $0$:
\begin{equation*}
P(\text{received bit 0}) = \sum_{ab \neq 11}  P(C_{0} \cap B_{00} \cap A_{ab}) = \frac{1}{2}.
\end{equation*} 
If Bob decides to always do nothing then this would be like a sticking strategy.  The probability that bit $0$ is received and Bob wins by sticking is $P(A_{00} \cap B_{00} \cap C_{0}) = 1/4$.  Hence we can compute the conditional probability:
\begin{equation}
P(\text{win if stick} \, | \, \text{received bit 0}) = \frac{1/4}{1/2} = \frac{1}{2}.
\end{equation}  
If an always switching strategy is adopted, then there are two possibilities ($01$ or $10$) each occuring with probability $1/2$.  In this case probability of winning if switched and bit $0$ is received is $P(A_{01} \cap B_{00} \cap C_{0} \cap D_{01}) + P(A_{10} \cap B_{00} \cap C_{0} \cap D_{10}) = 1/8$. With that we compute,
\begin{equation}
P(\text{win if switch} \, | \, \text{received bit 0}) = \frac{1/8}{1/2} = \frac{1}{4}. 
\end{equation}
It is an advantage to stick ie Bob should do nothing.  This strategy may used as an error-correcting design for reliability issues in practical quantum networks~\cite{simon2017towards, ren2017ground}

\textit{Conclusions:} We have reformulated PBR theorem into a Monty Hall game.  We argue that future investigation of Monty Hall concepts applied to antidistinguishability scenarios will lead to novel quantum protocols.

\textit{Acknowledgments:} 
DR was indirectly supported by the Marsden fund, administered by the Royal Society of New Zealand.
MV was directly supported by the Marsden fund, administered by the Royal Society of New Zealand.

         
\end{document}